\documentclass[10pt,prl,aps,twocolumn,floatfix,preprintnumbers,superscriptaddress,nofootinbib,nobibnotes]{revtex4-2}
 \usepackage{tikz,cancel}
   \usepackage[normalem]{ulem}
   \usetikzlibrary{shapes.geometric,arrows.meta,positioning}
\usepackage{amsfonts,amsthm,amssymb,physics}
\usepackage{stmaryrd}
\usepackage{mathtools}
\usepackage{lmodern}
\usepackage{CJKutf8,enumitem}
\usepackage[T1]{fontenc}
\usepackage[utf8]{inputenc}
\usepackage{graphicx}
\usepackage{comment}
\usepackage{bbm,soul}
\usepackage{mathrsfs}

\DeclareMathAlphabet{\mathbfi}{OML}{cmm}{b}{it}
\usepackage{tikz}
\definecolor{mLightBrown}{HTML}{EB811B}
\usetikzlibrary{shapes.geometric, arrows}
\tikzstyle{startstop} = [rectangle, rounded corners, 
minimum width=1cm, 
minimum height=1cm,
text centered, 
draw=black,
fill=pink!50!yellow!90!orange]

\tikzstyle{decision} = [diamond, 
minimum width=2.85cm, 
minimum height=2.92cm, 
text width=2cm, 
text centered, 
draw=teal!70!green, 
fill=teal!70!green!60]

\tikzstyle{process} = [rectangle, rounded corners, 
minimum width=1.5cm, 
minimum height=1.05cm, 
text centered, 
text width=1.8cm, 
draw=mLightBrown, 
fill=mLightBrown!40!white]

\tikzstyle{arrow} = [thick,black!70,-latex,>=stealth]

\definecolor{crimson}{rgb}{0.7, 0.08, 0.24}
\definecolor{green(munsell)}{rgb}{0.0, 0.7, 0.4}
\definecolor{Blu}{rgb}{0.0, 0.4, 0.7}
\definecolor{Ble}{rgb}{0.0,0.25,0.85}
\definecolor{darkmagenta}{rgb}{0.55, 0.0, 0.55}
\definecolor{purple(munsell)}{rgb}{0.62, 0.0, 0.77}

\usepackage[colorlinks=true,
  linkcolor=Ble,
  citecolor=crimson,
  urlcolor=Ble]{hyperref}
\usepackage{xcolor}

\newcommand{\rmd}{\ensuremath{\text{d}}}

\begin{document}

\title{A Proposal for Quantum Reference Frames in Quantum Field Theory and their Experimental Exploration}

\author{Ivan Agullo}
\email{agullo@valer-f.es}
\affiliation{Instituto de Física Corpuscular (IFIC), CSIC-Universitat de València and Departament de Física Teórica, Spain}
\affiliation{ValER, Valencian Foundation for Excellence in Research, Calle Mayor 83-1º-L9, 12001
Castelló de la Plana, Spain}
\affiliation{Department of Physics and Astronomy, Louisiana State University, Baton Rouge, LA 70803, USA}

\author{Fabio M. Mele} 
\email{fabio.mele@uv.es}
\affiliation{Instituto de Física Corpuscular (IFIC), CSIC-Universitat de València and Departament de Física Teórica, Spain}
\affiliation{Department of Physics and Astronomy, Louisiana State University, Baton Rouge, LA 70803, USA}

\date{\today}

\begin{abstract}
We construct quantum reference frames (QRFs) in quantum field theory, using localized modes of a scalar field as frames for its shift symmetry.~We import tools from relativistic quantum information to explicitly compute entropies and correlation measures of relational observables, investigating their frame dependence and the new features arising from the field-theoretic character of the system.~Unlike QRFs for systems with finitely many degrees of freedom, localized field-theoretic QRFs are necessarily mixed in physically relevant states and generically correlated with the subsystems they describe.~These aspects of QRFs for fields affect the properties of relational observables in a subtle way.~Furthermore, we put forward a proposal for testing these predictions with existing quantum simulators based on ultracold atoms and quantum fluids, thus connecting QRF physics with analogue gravity platforms.  
\end{abstract}

\maketitle

\section{Introduction}

The idea that reference frames are themselves quantum systems---quantum reference frames (QRFs)---has a long history:~it dates back to the 1960s~\cite{Aharonov:1967zza} and was further developed in the 1980s~\cite{Aharonov:1984}.~Over the past decades, QRFs have attracted renewed attention~\cite{Bartlett:2006tzx,angeloPhysicsQuantumReference2011a,Loveridge:2017pcv,Giacomini:2017zju,Vanrietvelde:2018pgb,Vanrietvelde:2018dit,Krumm:2020fws,Hoehn:2021flk,delaHamette:2021oex,Carette:2023wpz} and have been extended to new contexts with a wide range of applications, from quantum information and quantum foundations to gauge theory and quantum gravity (see, e.g.,~\cite{Smith:2017pwx,Giacomini:2018gxh,Castro-Ruiz:2019nnl,Hoehn:2019fsy,Hoehn:2020epv,Giacomini:2021gei,AliAhmad:2021adn,Carrozza:2021gju,Castro-Ruiz:2021vnq,delaHamette:2021piz,delaHamette:2021iwx,Carrozza:2022xut,Goeller:2022rsx,Kabel:2023jve,Hoehn:2023ehz,Suleymanov:2023wio,Kabel:2024lzr,Fewster:2024pur,DeVuyst:2024khu,Cepollaro:2024rss,Chataignier:2024eil,Giesel:2024xtb,DeVuyst:2024fxc,Carrozza:2024smc,Araujo-Regado:2024dpr,Araujo-Regado:2025ejs,delaHamette:2026rtv,Lacambra:2026twp,Rothlin:2026fqy,Luppi:2026qyv,Valente:2026ggh}).~Most of this work, however, has dealt with systems with finitely many degrees of freedom, often non-relativistic, and has remained theoretical.~Only recently have QRF ideas begun to be extended to quantum fields \cite{Hoehn:2023axh,Glowacki:2024jhi,DeVuyst:2024khu,DeVuyst:2024fxc,Fewster:2024pur,Fedida:2025viy,Fewster:2025ijg,Freidel:2025ous,Thiemann:2026vaj,Freidel:2026stu,Chen:2026kui,Aguilar-Gutierrez:2026svf}, with developments that have so far focused on structural and foundational aspects.~Two broad challenges thus stand out:~carrying the framework into quantum field theory, and endowing its formal structures with concrete, measurable physical content.~The present work addresses both, and takes a further step by proposing, to our knowledge, the first experimental realization of a quantum reference frame in a quantum-field simulator.

Our approach is deliberately bottom-up:~rather than beginning with a general, axiomatic framework for QRFs in quantum field theory (QFT), we start from a concrete theory---a free massless scalar field and its shift symmetry---and construct a specific family of QRFs within it.~Although this sacrifices generality at the outset, it keeps the construction tied to an experimentally accessible setup so that each mathematical ingredient acquires a clear physical interpretation and the analysis can be guided by physical intuition.~Our aim is thus to establish a concrete, physically grounded starting point that can subsequently be generalized to other field theories, and extended to accommodate both physical and gauge symmetries.~Further details will be given in a companion paper \cite{QRFQFTlong}.

We address the following questions:~(i) In the presence of a symmetry, how can a quantum field be described from within, without recourse to external reference frames, or even to other fields?~(ii) What are the distinctive features of QRFs in QFT, and how do they differ from their counterparts in systems with finitely many degrees of freedom?~(iii) Do key insights of the QRF program---such as the frame dependence of subsystems, correlations, and entanglement---extend to QFT?~(iv) Can experimentally accessible signatures of QRF physics be identified?~We use units $\hbar=c=1$, and Boltzmann constant $k_B=1$, except in the last section, where we restore $\hbar$ and the role of $c$ is played by the speed of sound $c_{\rm s}$.

\section{The theory}

We consider a real, massless scalar field on a $(D+1)$-dimensional spacetime, obeying the Klein--Gordon equation $\Box\hat\Phi(x)=0$.~For this linear theory, the field algebra can be formulated equivalently in a canonical or in a manifestly covariant framework (see, e.g., \cite{Wald:1995yp}).~In the canonical formulation, one starts from the field $\hat\Phi(\vec x)$ and its conjugate momentum $\hat\Pi(\vec x)$, defined on a Cauchy hypersurface $\Sigma$.~We adopt this formulation here because it makes the connection with the experimental implementation more transparent.

Recall that the pointwise operators $\hat\Phi(\vec x)$ and $\hat\Pi(\vec x)$ are not well-defined;~well-defined operators are obtained by smearing them against suitable test functions.~For the purposes of this article, we take the test functions to belong to $C_0^\infty(\Sigma)$, the space of real-valued smooth functions of compact support on $\Sigma$, and define the smeared field operators
\begin{equation}\label{smearedopts}
\hat\Phi_f:=\int_\Sigma\rmd^DV\,f(\vec x)\,\hat\Phi(\vec x)\,,\  
\hat\Pi_g:=\int_\Sigma\rmd^Dx\,g(\vec x)\,\hat\Pi(\vec x)\,,
\end{equation}
with $d^DV$ the volume element in $\Sigma$.~These operators, together with the identity $\hat{\mathbb{I}}$, generate the canonical algebra $\mathcal{A}$:~its elements are finite linear combinations of products of generators, with the $\ast$-relations $\hat\Phi_f^*=\hat\Phi_f$ and $\hat\Pi_g^*=\hat\Pi_g$, and the canonical commutation relations $[\hat\Phi_f,\hat\Pi_g]=i\int_\Sigma\rmd^DV\,f(\vec x)\,g(\vec x)\,\hat{\mathbb{I}}$. The domain issues that arise when these unbounded operators are represented on a Hilbert space can be avoided by working instead with the Weyl algebra.
 
\subsection{Localized subsystems}

Motivated by experimental accessibility, we focus on subsystems that are spatially localized within compact regions and contain finitely many degrees of freedom.~As a starting point, consider a subsystem containing a single degree of freedom, which we refer to as a {\em single-mode subsystem}.~Let $f_A$ and $g_A$ be two functions in $C_0^\infty(\Sigma)$ satisfying (i) ${\rm supp}(f_A),\,{\rm supp}(g_A)\subseteq I_A$, for some compact region $I_A\subset\Sigma$, and (ii) $\int_\Sigma\rmd^DV\,f_A(\vec x)\,g_A(\vec x)\neq 0$.~Condition (i) ensures localization within $I_A$, while condition (ii) ensures that $\hat\Phi_{f_A}$ and $\hat\Pi_{g_A}$ do not commute and, after a simple rescaling, form a canonically conjugate pair.~The algebra $\mathcal{A}_A$ generated by $\hat\Phi_{f_A}$ and $\hat\Pi_{g_A}$ defines a single-mode subsystem localized within $I_A$;~it is isomorphic to the canonical algebra of a single degree of freedom, e.g.~a harmonic oscillator.~This construction does not exhaust the most general class of single-mode subsystems (see, e.g.,~\cite{QRFQFTlong,Agullo:2023fnp,Agullo:2024har}), but it suffices for our purposes.

An $N$-mode subsystem is  obtained as a collection of $N$ mutually commuting single-mode subsystems, i.e., with $\int_\Sigma\rmd^DV\,f_i(\vec x)\,g_j(\vec x)=0$ for $i\neq j$.

A feature of QFT which sharply contrasts standard quantum mechanical systems is that a localized mode of a QFT can never be found in a pure quantum state when the field is prepared in any state of bounded energy \cite{Hollands:2017dov}.~This is because, by Reeh-Schlieder theorem \cite{Reeh:1961ujh}, all localized modes are necessarily correlated and entangled with modes supported in other regions \cite{Hollands:2017dov,Verch:2004vj};~hence the von-Neumann entropy of localized subsystems cannot vanish.

It follows that, in QFT, it is natural to characterize localized subsystems by their operator algebras, rather than by a Hilbert space associated with the subsystem, since vectors in such Hilbert space would describe pure states of the subsystem, which are not physically realizable.~This represents a departure from some of the existing approaches to QRFs~\cite{Bartlett:2006tzx,Giacomini:2017zju,delaHamette:2021oex,Castro-Ruiz:2021vnq}, which have primarily been formulated for systems with finitely many degrees of freedom and typically characterize the configurations of the frame using pure states (see however \cite{Bojowald:2019mas,Bojowald:2022caa,Bojowald:2022jem,DeVuyst:2025ezt} for an algebraic approach to QRFs in constrained systems).~In this respect, we find the mathematical and conceptual machinery employed, e.g., in the operational approach to QRFs \cite{Busch:2016rui,Loveridge:2016tnh,Loveridge:2017pcv,Carette:2023wpz}  well suited to our field-theoretic setting.

\subsection{The symmetry}

The scalar field theory under consideration possesses a shift symmetry,
\begin{equation}
\hat\Phi(\vec x)\longmapsto\hat\Phi(\vec x)+\lambda\,\hat{\mathbb{I}}\,,\qquad
\hat\Pi(\vec x)\longmapsto\hat\Pi(\vec x)\,,
\end{equation}
with $\lambda\in\mathbb{R}$.~This transformation acts on field space rather than on spacetime.~It is of interest because it provides a simple analogue of gauge transformations in electrodynamics, $\hat A\mapsto\hat A+\rmd\beta\,\hat{\mathbb{I}}$, and therefore provides a convenient setting in which to develop tools that can  be extended to them.~Note, however, that for the scalar field the shift is a genuine symmetry rather than a gauge redundancy.

The action of the symmetry group on the operator algebra $\mathcal{A}$ follows directly from the transformation of the operator-valued distributions $\hat\Phi(\vec x)$ and $\hat\Pi(\vec x)$:
\begin{equation}
\alpha_\lambda\big[\hat\Phi_f\big]=\hat\Phi_f+\lambda\,\bar f\,\hat{\mathbb{I}}\,,\qquad
\alpha_\lambda\big[\hat\Pi_g\big]=\hat\Pi_g\,,
\end{equation}
for all $f,g\in C_0^\infty(\Sigma)$, where $\bar f:=\int_\Sigma\rmd^DV\,f(\vec x)$ is the integral of $f$.~Thus, all smeared momentum operators are invariant under the shift, whereas a smeared field operator $\hat\Phi_f$ is invariant if and only if $\bar f=0$.

\subsection{The algebra of observables}

In the absence of an external reference frame defining an origin for the values of $\hat\Phi$, physical observables must be represented by shift-invariant operators (see~\cite{Loveridge:2017pcv} for a broader discussion of the relation between symmetries and observables).~The algebra of observables is therefore the subalgebra $\mathcal{A}_{\rm obs}\subset\mathcal{A}$ of all invariant elements, $\alpha_\lambda[\hat O]=\hat O$ for all $\lambda\in\mathbb{R}$.~It is generated by the identity, all smeared momenta $\hat\Pi_g$, and the smeared fields $\hat\Phi_f$ with $\bar f=0$.

Notice that there is no need to restrict the states to be invariant:~restricting the algebra of observables is sufficient.~Indeed, the restriction to $\mathcal{A}_{\rm obs}$ induces an equivalence relation on the space of states:~two states are physically equivalent whenever they assign the same expectation values to all elements of $\mathcal{A}_{\rm obs}$, since no measurement can distinguish them.~Each equivalence class contains a shift-invariant state, obtained, formally, by averaging any of its elements over the symmetry group.~However, because the symmetry group is non-compact, such invariant states are not regular (in particular, they cannot be represented by density matrices in the Fock representation) and must be treated with care (much as no normalizable state of a free particle is invariant under spatial translations).~Restricting the algebra rather than the states avoids such idealized states altogether.\footnote{The situation is different for gauge symmetries, where the gauge redundancy forces an additional restriction on the physical states (the generators of gauge transformations must vanish on them) which cannot be achieved by restricting the algebra alone~\cite{Vanrietvelde:2018pgb} (see also \cite{Hoehn:2023ehz},~Sec.~II B 3).}

\subsection{Measurements}

An important ingredient of measurement theory is the specification of how a physical system can be probed.~Realistic measurements are necessarily imperfect, owing to finite resolution, dark counts, and other sources of uncertainty.~In quantum theory, such measurements are described by positive operator-valued measures (POVMs)~\cite{Busch:1997book}.

We are interested in POVMs associated with the smeared field operator $\hat\Phi_f$, for an arbitrary $f\in C_0^\infty(\Sigma)$.~For concreteness, we consider a {\em Gaussian-modulated POVM}~\cite{Mandrysch:2024vwl}, defined by the family of positive operators $\{\hat E_f(u)\}_{u\in\mathbb{R}}$,
\begin{equation}\label{GMPOVM}
\hat E_f(u)=\frac{1}{\sqrt{2\pi}\epsilon}\exp\left[-\frac{\big(\hat\Phi_f-u\big)^2}{2\epsilon^2}\right]\,,\,
\int_{\mathbb{R}}\rmd u\,\hat E_f(u)=\hat{\mathbb{I}}\,,
\end{equation}
where $\epsilon>0$ characterizes the measurement resolution and may depend on $f$.~Strictly speaking, $\hat E_f(u)$ is an (operator-valued) density:~the effects of the POVM are $\hat E_f(\Delta)=\int_\Delta\rmd u\,\hat E_f(u)$, and the probability that the outcome lies in an interval $\Delta\subset\mathbb{R}$ (more precisely, a Borel set of $\mathbb{R}$) is equal to the expectation value of $\hat E_f(\Delta)$.

Physically, a POVM specifies an experimental question, its effects corresponding to the possible outcomes.~Here, the question is \emph{what is the value of $\hat\Phi_f$?}, and $\hat E_f(u)\,\rmd u$ represents the outcome that $\hat\Phi_f$ is found in the vicinity of $u$, within the resolution $\epsilon$. 

The mathematical properties of this POVM have been studied in~\cite{Mandrysch:2024vwl} (see also \cite{Jubb:2021mvn,Oeckl:2024diw} for complementary discussions).~Its most important features are: (i) as $\epsilon\to0$, it approaches the textbook projective measurement of $\hat\Phi_f$; (ii) it transforms covariantly under the shift symmetry, $\alpha_\lambda\big[\hat E_f(u)\big]=\hat E_f(u-\lambda\bar f)$; (iii) it can be implemented within the Fewster--Verch measurement scheme for QFT~\cite{Fewster:2018qbm,Fewster:2023cfq}; and (iv) it admits a movable Heisenberg cut (i.e., the same POVM is obtained regardless of where the boundary between the quantum description and the classical apparatus is placed~\cite{Mandrysch:2024vwl}).~Property (ii) is precisely what the operational approach to QRFs requires of a ``frame observable''~\cite{Loveridge:2017pcv,Carette:2023wpz}.\footnote{See also \cite{Ludescher:2026xsw} for a recent discussion of general covariant POVMs within the perspective-neutral approach to QRFs.} Moreover, its Gaussian form makes it particularly well suited to analytical calculations.
 
\section{Localized QRF\texorpdfstring{\lowercase{s}}{s}}

In our setup, a localized QRF consists of two ingredients: (i) a localized subsystem of the field, and (ii) a frame POVM, i.e., a POVM on that subsystem that is covariant under the symmetry and specifies our capability to extract information from it.

For the frame subsystem, we take a localized single-mode subsystem, with algebra $\mathcal{A}_R$ generated by $\hat\Phi_{f_R}$ and $\hat\Pi_{g_R}$, where $f_R,g_R\in C_0^\infty(\Sigma)$ are supported in a compact region $I_R$ and satisfy $\int_\Sigma\rmd^DV\,f_R(\vec x)\,g_R(\vec x)\neq0$.~We further require $\bar f_R\neq0$;~otherwise $\hat\Phi_{f_R}$---and hence the entire algebra $\mathcal{A}_R$---would be invariant under the shift symmetry, and   the frame would be insensitive to the symmetry and could not be used to resolve the ambiguity it introduces.

For the frame POVM, we take the Gaussian-modulated POVM~\eqref{GMPOVM} with $f=f_R$ and resolution $\epsilon_R$. In the terminology of~\cite{Carette:2023wpz}, this QRF is unsharp, and hence non-ideal, since $\epsilon_R>0$; it is complete, since the shift group acts freely on the space of outcomes, $u\mapsto u+\lambda\bar f_R$; and non-localizable in the operational sense \cite{Carette:2023wpz} (not to be confused with the spatial localization of the frame) since, for any finite resolution $\epsilon_R>0$, the nonzero tails of the Gaussian response prevent the probability of obtaining an outcome in any bounded interval from becoming arbitrarily close to unity.

\section{Relational observables}

Let $\hat O\in\mathcal{A}$ be an operator commuting with the frame algebra $\mathcal{A}_R$, not necessarily invariant under the shift symmetry.~Our goal is to use the frame to construct from $\hat O$ a shift-invariant observable belonging to $\mathcal{A}_{\rm obs}$.~This is achieved by means of the \emph{relativization map}~\cite{Loveridge:2017pcv}, defined by
\begin{equation}\label{relmap}
\hat F_{\hat O|R}(u)=|\bar f_R|\int_{\mathbb{R}}\rmd\lambda\;\alpha_\lambda\big[\hat E_{f_R}(u)\big]\,\alpha_\lambda\big[\hat O\big]\,.
\end{equation}
The operator $\hat F_{\hat O|R}(u)$ is interpreted as the observable describing $\hat O$ relative to the reference frame, conditional on the frame being found in the configuration $u$.~It therefore depends explicitly on the POVM used to probe the frame. 

The relativization map can be understood as an averaging of the product $\hat E_{f_R}(u)\,\hat O$ over the symmetry group. The resulting operator is automatically invariant under the group action, and hence belongs to $\mathcal{A}_{\rm obs}$. Although the group is non-compact, the integral converges, since $|\bar f_R|\, \int\rmd\lambda\,\alpha_\lambda[\hat E_{f_R}(u)]=\hat{\mathbb{I}}$; the prefactor $|\bar f_R|$ ensures that $\hat F_{\hat{\mathbb{I}}|R}(u)=\hat{\mathbb{I}}$, so that the map sends POVMs to POVMs. Related group-averaging constructions have a long history in gauge theory and quantum gravity~\cite{Rovelli:2004tv,Thiemann:2007pyv,Tambornino:2011vg,Rovelli:1990ph,Rovelli:1990pi,Dittrich:2004cb,Dittrich:2005kc}.~(See also \cite{Hoehn:2019fsy,Krumm:2020fws,Chataignier:2020fys,AliAhmad:2021adn,delaHamette:2021oex,Hoehn:2023ehz,Chataignier:2024eil} for related discussions in the context of QRFs.)

Additional intuition can be obtained from concrete examples.~For the Gaussian-modulated POVM introduced above, the integral in Eq.~\eqref{relmap} can be evaluated analytically.~In particular, consider the Weyl operator $\hat W_{f,g}=\exp\big[i\big(\hat\Phi_f+\hat\Pi_g\big)\big]$, with $\int_\Sigma\rmd^DV\,f\,g_R=\int_\Sigma\rmd^DV\,f_R\,g=0$, so that $W_{f,g}$ commutes with the frame algebra $\mathcal{A}_R$.~One finds
\begin{equation}\label{relWeyl}
\hat F_{\hat W_{f,g}|R}(u)=e^{-\frac{\epsilon_R^2}{2}\frac{\bar f^2}{\bar f_R^2}}\;
e^{i\left(\hat\Phi_f+\hat\Pi_g-\frac{\bar f}{\bar f_R}\hat\Phi_{f_R}+u\,\frac{\bar f}{\bar f_R}\,\hat{\mathbb{I}}\right)},
\end{equation}
where $\epsilon_R$ is the resolution of the frame POVM. Derivatives of this expression yield the relativized versions of (symmetrically ordered) polynomials in $\hat\Phi_f$ and $\hat\Pi_g$. For a smeared momentum operator, one readily obtains $\hat F_{\hat\Pi_g|R}(u)=\hat\Pi_g$, as expected, since $\hat\Pi_g$ is already shift-invariant.~For the smeared field operator, on the other hand,
\begin{equation}\label{eq:relfield}
\hat F_{\hat\Phi_f|R}(u)=\hat\Phi_f-\frac{\bar f}{\bar f_R}\hat\Phi_{f_R}+u\,\frac{\bar f}{\bar f_R}\,\hat{\mathbb{I}}\,.
\end{equation}
The interpretation is straightforward.~The last term, proportional to the identity, is an offset that records the configuration $u$ of the frame.~The first two terms combine into $\hat\Phi_{f_{\rm rel}}$, with $f_{\rm rel}:=f-\frac{\bar f}{\bar f_R}f_R$, and represent the value of $\hat\Phi_f$ relative to $\hat\Phi_{f_R}$;~the factor $\bar f/\bar f_R$ ensures that $\bar f_{\rm rel}=0$, hence $\hat\Phi_{f_{\rm rel}}$ is shift-invariant.

In essence, the relativization map acting on a smeared field $\hat\Phi_f$ replaces $f$ by $f_{\rm rel}$, a test function with zero integral.~The QRF framework, however, does more than select a class of mode functions:~it endows them with an operational meaning by tying them to a physical subsystem of the field---the frame---whose fluctuations and correlations with the system enter the predictions, as we show below.~Furthermore, our capability to probe the frame, encoded in the frame POVM, is built into the relational observables and into the algebra they form.

Similarly, applying the relativization map to the Gaussian-modulated POVM of $\hat\Phi_f$, with resolution $\epsilon_f$, yields a relational POVM.~Using Eq.~\eqref{relWeyl}, one finds that it is again Gaussian-modulated, now for the relational observable $\hat F_{\hat\Phi_f|R}(u)$, with resolution $\epsilon_{\rm eff}^2=\epsilon_f^2+\Big(\frac{\bar f}{\bar f_R}\Big)^2\epsilon_R^2$.

Note that the relativization map is linear but not multiplicative:~for instance, $\hat F_{\hat\Phi_f^2|R}(u)=\big[\hat F_{\hat\Phi_f|R}(u)\big]^2+\big(\bar f/\bar f_R\big)^2\epsilon_R^2\,\hat{\mathbb{I}}$, the last term reflecting the unsharpness of the frame.~This is compatible with the observation in \cite{Carette:2023wpz} that, for general POVMs, the image of the relativization map does not form an algebra, but an operator space.~For the Gaussian-modulated POVM written in \eqref{GMPOVM}, however, the image of $\hat{F}_{\bullet|R}(u)$ can be endowed with a deformed product, which depends on $\epsilon_R$ and the smearing functions, with respect to which it forms a $*$-algebra and the relativization map becomes a $*$-homomorphism.~For the smeared field operators, the deformed product reads
\begin{equation}\label{eq:star}
    \hat F_{\hat \Phi_f|R}(u)\star_{\epsilon_R}\hat F_{\hat \Phi_{f'}|R}(u)=\hat F_{\hat \Phi_f|R}(u)\hat F_{\hat \Phi_{f'}|R}(u)+\epsilon_R^2\,\frac{\bar f\bar f'}{\bar f_R^2}\,\hat{\mathbb I}\;,
\end{equation}
so that $\hat{F}_{\Phi_{f}\Phi_{f'}|R}(u)=\hat F_{\hat \Phi_f|R}(u)\star_{\epsilon_R}\hat F_{\hat \Phi_{f'}|R}(u)$.~Note that Eq.~\eqref{eq:star} reduces to the ordinary product in the limit $\epsilon_R\to0$, or whenever either of the two field operators is shift-invariant.~Products involving relativized smeared momentum operators are unaffected by the deformation.~We refer to \cite{QRFQFTlong} for further details.

Given a state $\omega$ of the field, Eq.~\eqref{relWeyl} provides a direct means of computing moments and correlation functions of relational observables, where products are understood in the deformed algebra.~It is instructive to consider the variance of the relational field operator $\hat\Phi_{f_{\rm rel}}$ (corresponding to $\hat{F}_{\hat{\Phi}_f|R}(u)$ after omitting  the offset proportional to $u$, since it contributes in a trivial way):
\begin{equation}
\begin{aligned}
{\rm var}_{\star_{\epsilon_R}}\big(\hat\Phi_{f_{\rm rel}}\big)=&\;{\rm var}\big(\hat\Phi_f\big)+\left(\frac{\bar f}{\bar f_R}\right)^2{\rm var}\big(\hat\Phi_{f_R}\big)+\left(\frac{\bar f}{\bar f_R}\right)^2\epsilon_R^2\\
&-2\,\frac{\bar f}{\bar f_R}\,{\rm cov}\big(\hat\Phi_f,\hat\Phi_{f_R}\big)\,,
\end{aligned}
\end{equation}
where ${\rm cov}(\hat A,\hat B):=\langle \hat A\hat B\rangle -\langle \hat A\rangle \,\langle \hat B\rangle $.~(In $1+1$ dimensions, and at finite temperature in $2+1$ dimensions, the individual terms on the right-hand side are IR-divergent, but their sum is finite, since $\bar f_{\rm rel}=0$.)~Interestingly, the variance of the relational field observable $\hat\Phi_{f_{\rm rel}}$ depends on the intrinsic variances of the system and frame operators, on the unsharpness of the frame through $\epsilon_R$, and also on the correlations between them.~Such correlations are ubiquitous in QFT and depend, among other factors, on the spatial separation between the supports of $f$ and $f_R$.~Consequently, in QFT, the fluctuations of a system observable relative to a QRF generically depend on the location of the frame.

\section{Relational subsystems}

Let $\mathcal{A}_S$ and $\mathcal{A}_R$ be the single-mode subalgebras associated with the system and the frame, respectively, which we assume to commute (as, e.g., when $S$ and $R$ are spatially separated).~We define the relational algebra $\mathcal{A}_{S|R}(u)$ relative to the frame $R$ in configuration $u$ as the algebra generated by the relational operators $\hat F_{\hat\Phi_{f_S}|R}(u)$ and $\hat F_{\hat\Pi_{g_S}|R}(u)$, together with the identity operator, with the $\star_{\epsilon_R}$-product.~Importantly, the relativization map \eqref{eq:relfield}, or equivalently \eqref{relWeyl} for the Weyl operators, preserves the canonical commutation relations with respect to the deformed product.~Consequently, $\mathcal{A}_{S|R}(u)$ defines a \emph{relational single-mode subsystem} generated by the conjugate pair $\hat\Phi_{f_{\rm rel}}+u\,\frac{\bar f_S}{\bar f_R}\,\hat{\mathbb{I}}$, $\hat\Pi_{g_S}$.

Because $\mathcal{A}_{S|R}(u)$ involves degrees of freedom of the frame, it differs from $\mathcal{A}_S$, and two distinct frames generally define different relational subsystems.~This provides a field-theoretic manifestation of the QRF-dependence of subsystems, previously studied for systems with finitely many degrees of freedom~\cite{AliAhmad:2021adn,delaHamette:2021oex,Castro-Ruiz:2021vnq,Hoehn:2023ehz}.~In QFT, the relational subsystem also changes if we vary the separation between system and frame.

Relational observables with respect to different frames are related by a simple transformation law:~for two frames $R_1$ and $R_2$, and, for simplicity, vanishing offsets,
\begin{equation}
\hat F_{\hat\Phi_{f_S}|R_2}(0)=\hat F_{\hat\Phi_{f_S}|R_1}(0)-\frac{\bar f_S}{\bar f_{R_2}}\,\hat F_{\hat\Phi_{f_{R_2}}|R_1}(0)\,,
\end{equation}
(the generalization to non-zero off-sets is straightforward) while relational momenta are unchanged.~This transformation can be implemented by a shift of $R_1$ conditioned on the relative field configuration of the frames:
\begin{equation}\label{activeQRFtransf}
\mathcal V_{R_1\to R_2}[\,\cdot\,]:=\int_{\mathbb R}\rmd P_{R_2|R_1}(q)\,\alpha^{R_1}_{q}[\,\cdot\,]\;,
\end{equation}
where $\rmd P_{R_2|R_1}(q)$ denotes the spectral measure of the relative field operator $\hat F_{\hat\Phi_{f_{R_2}}|R_1}(0)/\bar f_{R_2}$, and the super-script in $\alpha^{R_1}_{q}$ indicates shifts acting on the subsystem $R_1$ alone.~Eq.~\eqref{activeQRFtransf} resembles the \emph{active frame change transformations} (change of relational observables) introduced in \cite{delaHamette:2021oex} (cf.~Eq.~(120)).

When restricted to the relational subsystem $\mathcal A_{S|R_1}(0)$, the transformation \eqref{activeQRFtransf} defines a $*$-isomorphism onto $\mathcal A_{S|R_2}(0)$, with respect to their deformed products, if and only if the two frames resolve the shift parameter equally well, i.e.,~$\epsilon_{R_1}/|\bar f_{R_1}|=\epsilon_{R_2}/|\bar f_{R_2}|$.~Otherwise, changing to the less precise frame requires adding classical noise, and the converse change cannot be implemented by any physical operation, in line with results of \cite{Carette:2023wpz}.

\subsection{Entropy of relational subsystems}

We define the entropy $S_{S|R}$ of the relational subsystem $S|R$ as the von Neumann entropy of the restriction of the state $\omega$ of the field to the algebra $\mathcal{A}_{S|R}(u)$.~Since the relational subsystem depends on the frame, so, in general, does its entropy.~It depends on the frame only through $f_R$ (not through $g_R$), and through the frame's POVM.

\begin{figure}[t!]
    \includegraphics[width=\columnwidth]{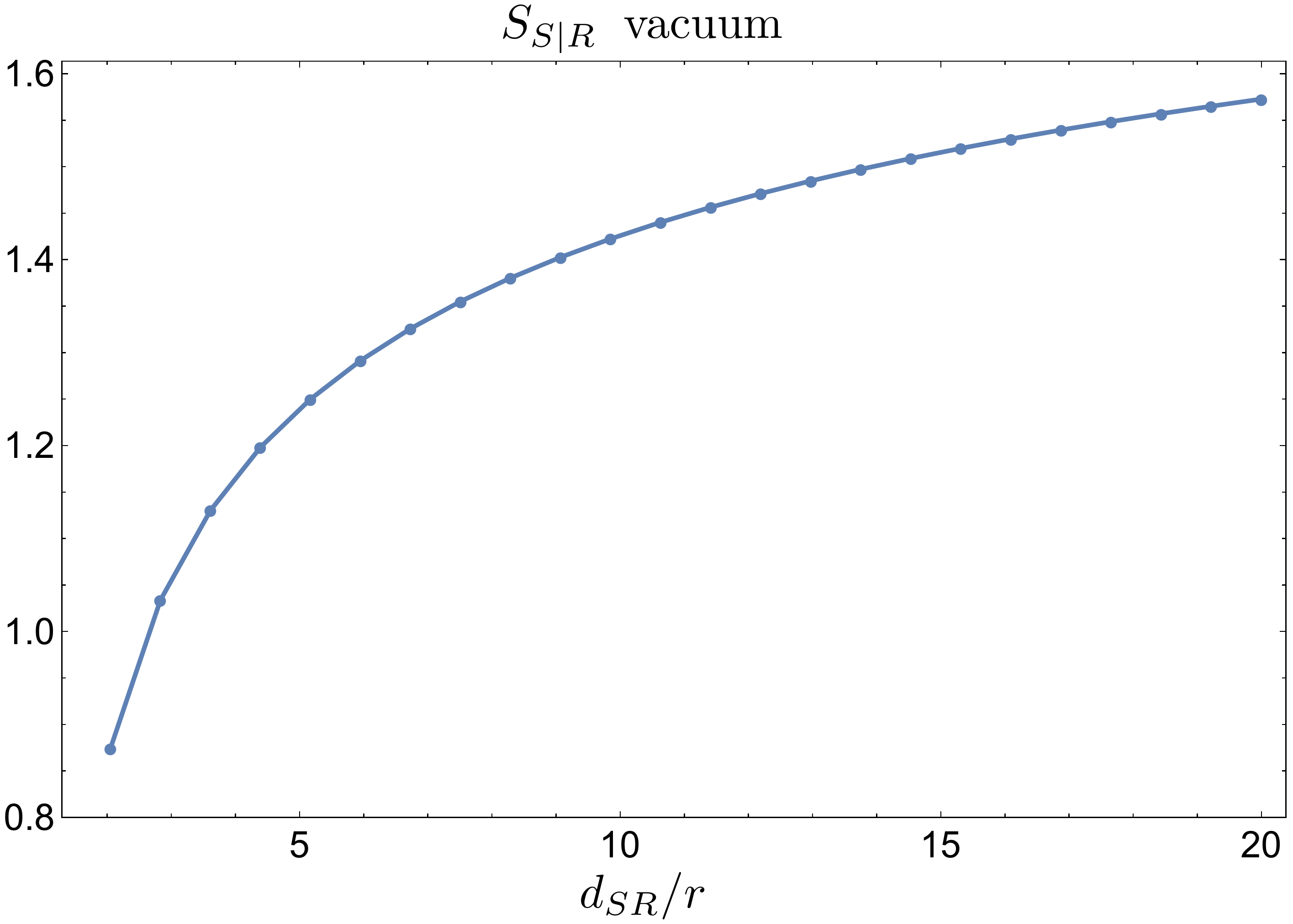}
    \caption{Entropy $S_{S|R}$ of the relational subsystem versus the separation $d_{SR}$ between system and frame.~The entropy grows with $d_{SR}$:~as correlations between system and frame weaken with separation, the relational subsystem becomes more mixed.~We have used $\epsilon_R\simeq 0.019$, which corresponds to a frame resolution of $2\%$, relative to the standard deviation of $\hat\Phi_{f_{\rm rel}}$ in vacuum (computed at the minimum separation within the  range considered).~The qualitative form of this plot is the same for other values of $\epsilon_R$.}
    \label{Fig:relent}
\end{figure}

To illustrate this effect, we consider a massless scalar field in flat $(1+1)$-dimensional spacetime, in the vacuum state.~Although this state is not defined on the full algebra $\mathcal{A}$, owing to infrared divergences, it is well defined on the algebra of observables $\mathcal{A}_{\rm obs}$, which is all that relational quantities require.~We define the system and frame modes by the compactly supported smearing functions~\cite{Agullo:2023fnp,Ribes-Metidieri:2025nfw}
\begin{equation}\label{eq:smearings}
f_j(x)=g_j(x)=N\left(1-\frac{(x-x_j)^2}{r^2}\right)\Theta\big(r-|x-x_j|\big),
\end{equation}
$j\in\{S,R\}$, where $r$ is the radius of their supports and $N$ is fixed by $\int\rmd x\,f_i\,g_j=\delta_{ij}$, so that $[\hat\Phi_{f_i},\hat\Pi_{g_j}]=i\,\delta_{ij}\,\hat{\mathbb{I}}$.~These functions are not smooth at the boundary of their supports, but are continuous, which suffices for all quantities below to be well-defined \cite{Agullo:2023fnp}.~Fig.~\ref{Fig:relent} shows the entropy $S_{S|R}$ as a function of $d_{SR}$, the separation between the centers $x_S$ and $x_R$ of the supports of the system and frame smearing functions.~This behavior suggests that correlations between the system and the frame can partially purify the relational subsystem, thereby reducing its entropy (this interpretation must be understood in light of the frame dependence of the subsystem itself).
\newpage

\subsection{Correlations between relational subsystems}

\begin{figure}[t!]
    \includegraphics[width=\columnwidth]{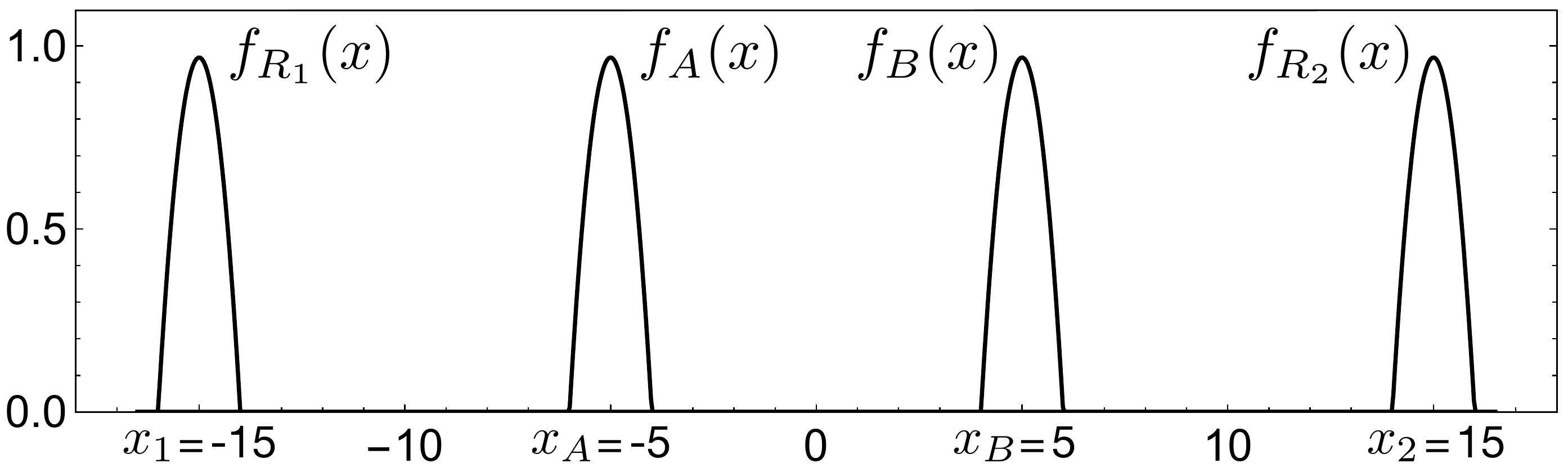}
    \caption{Setup for the computation of the correlations between relational subsystems:~two single-mode subsystems, $A$ and $B$, and two single-mode QRFs, $R_1$ and $R_2$, defined by the same smearing functions as in Eq.~\eqref{eq:smearings}, centered at $x_A$, $x_B$, $x_1$, and $x_2$ (measured in units of $r$) respectively, with pairwise disjoint supports.}
    \label{Fig:smearings}
\end{figure}

The correlations between relational subsystems reveal even more strikingly the richness that QFT brings to the physics of QRFs.~We consider the configuration shown in Fig.~\ref{Fig:smearings}, consisting of two subsystems, $A$ and $B$, and two reference frames, $R_1$ and $R_2$, defined by the same smearing functions as in the previous example, centered at $x_A$, $x_B$, $x_1$, and $x_2$, respectively, with pairwise disjoint supports.~Our goal is to investigate the frame dependence of correlations between the relational subsystems $A|R_i$ and $B|R_i$, for $i=1,2$.
\begin{figure}[t!]
    \includegraphics[width=\columnwidth]{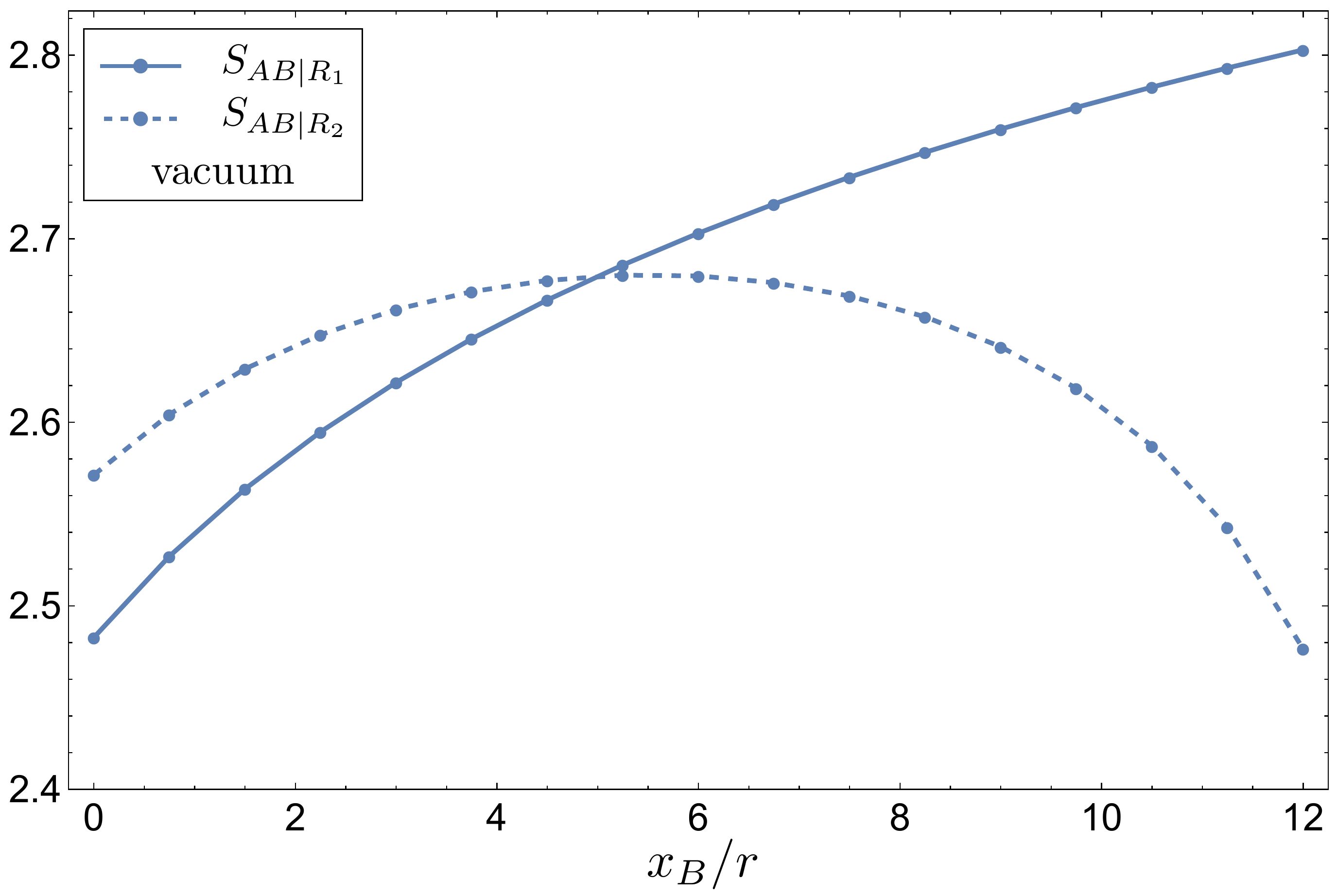}
    \caption{Entropy of the two-mode system $AB$ relative to the frames $R_1$ and $R_2$ versus the position $x_B$ of the center of subsystem $B$.~The entropy $S_{AB|R_1}$ varies monotonically with $x_B$, whereas $S_{AB|R_2}$ does not.~For concreteness, we have used the same value of $\epsilon_R$ as in Fig.~\ref{Fig:relent}, i.e.,~$\epsilon_R\simeq 0.019$.~The qualitative form of this plot is the same for other values of $\epsilon_R$.}
    \label{Fig:ABentropy}
\end{figure}

We first plot, in Fig.~\ref{Fig:ABentropy}, the entropy in vacuum of the joint system $AB$ relative to each frame, $R_1$ and $R_2$, as a function of the position $x_B$ of subsystem $B$.~The two behave qualitatively differently: $S_{AB|R_1}$ varies monotonically, whereas $S_{AB|R_2}$ does not.~This difference can be interpreted in terms of the correlations between the individual subsystems.~As $x_B$ increases, subsystem $B$ moves away from both $R_1$ and $A$, so that its correlations with both decrease monotonically, and $S_{AB|R_1}$ increases accordingly.~By contrast, as $x_B$ increases, $B$ approaches $R_2$ while moving away from $A$, and $S_{AB|R_2}$ reflects a competition between the decreasing correlations of $B$ with $A$ and the increasing correlations of $B$ with $R_2$.~The non-monotonic profile of $S_{AB|R_2}$, with a maximum, then follows from the fact that correlations between any two subsystems vary more rapidly when the subsystems are closer together.

\begin{figure}[t!]
    \includegraphics[width=\columnwidth]{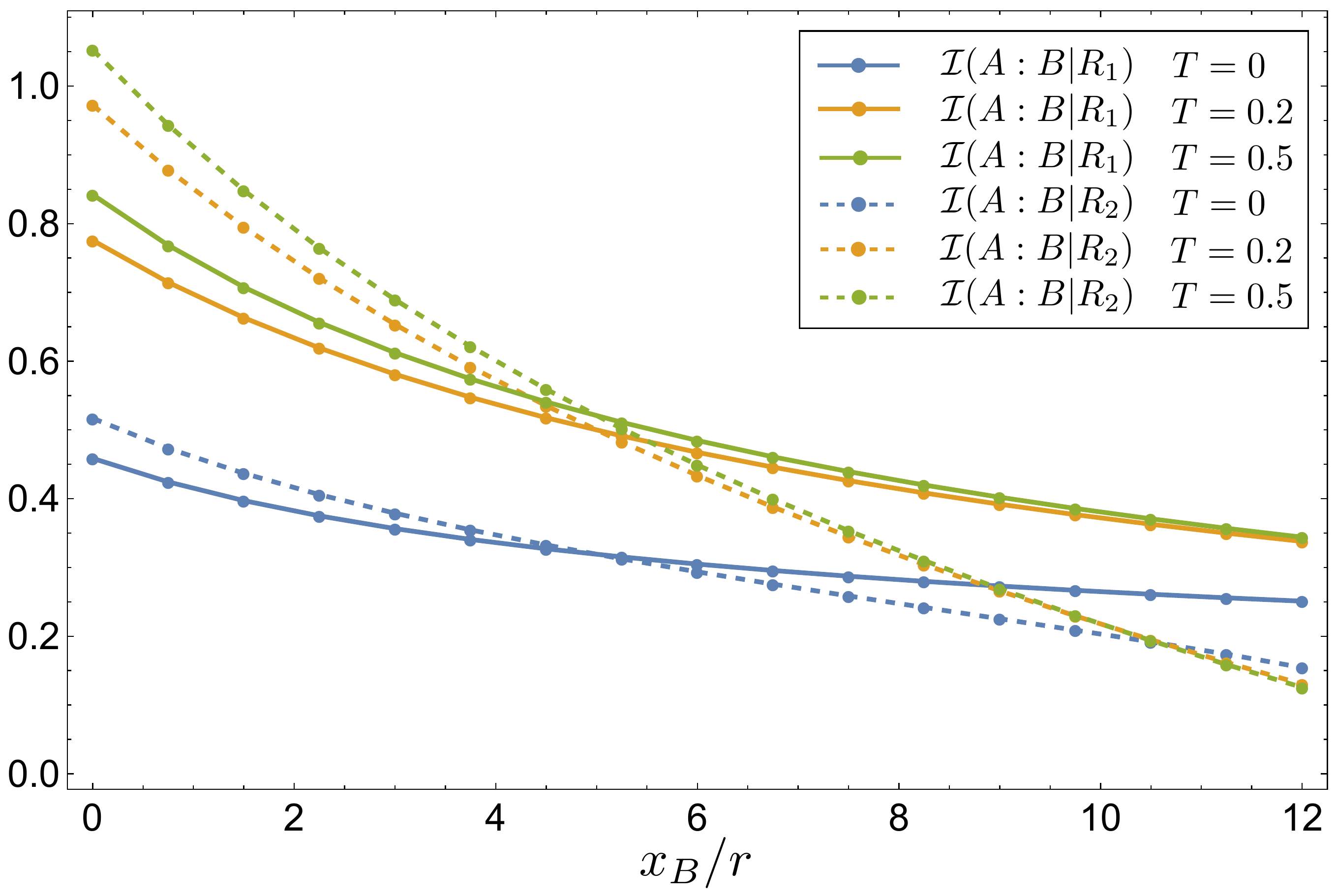}
    \caption{Mutual information between the relational subsystems $A|R_i$ and $B|R_i$, for $i=1,2$, versus the position $x_B$ of the center of subsystem $B$, at several temperatures (measured in units of $r$).~For large separations ($x_B/r\gtrsim10.5$), $\mathcal I(A{:}B|R_1)$ grows with temperature, whereas $\mathcal I(A{:}B|R_2)$ decreases.~The same value of $\epsilon_R$ as in the other plots is used ($\epsilon_R\simeq 0.019$).~The qualitative form of this plot is the same for other values of $\epsilon_R$.}
    \label{Fig:ABmutualinfo}
\end{figure}

Next, we investigate the mutual information (MI) between the relational subsystems $A|R_i$ and $B|R_i$, in the same configuration.~The MI, $\mathcal{I}(A{:}B|R_i):=S_{A|R_i}+S_{B|R_i}-S_{AB|R_i}$, quantifies the total---classical and quantum---correlations between them.~Fig.~\ref{Fig:ABmutualinfo} shows $\mathcal{I}(A{:}B|R_i)$ as a function of $x_B$, for a thermal state of the field at several temperatures.~Not only does the dependence of the MI on $x_B$ differ between the two frames, so does its response to temperature:~in our example, for $x_B/r\gtrsim10.5$, $\mathcal I(A{:}B|R_1)$ increases with temperature, whereas $\mathcal I(A{:}B|R_2)$ decreases.

The entropies and correlations of relational subsystems depend on multiple factors, such as the choice of smearing functions, the spatial separations between subsystems, and the state of the field.~Although analytical expressions can be derived for many specific configurations, they are often lengthy and offer limited additional insight.~The purpose of the preceding examples is instead to illustrate how the ubiquitous correlations of QFT enrich the physics of subsystem relativity, revealing features that go beyond those encountered in the systems with finitely many degrees of freedom studied to date~\cite{Giacomini:2017zju,Vanrietvelde:2018pgb,delaHamette:2020dyi,AliAhmad:2021adn,delaHamette:2021oex,Castro-Ruiz:2021vnq,Hoehn:2023ehz}.

We have complemented this analysis with other smearing functions, and by computing entanglement, quantified by logarithmic negativity~\cite{VidalNegativity,Plenio2005logneg,Audenaert2003}, finding qualitatively similar behaviors.~These results will be reported elsewhere~\cite{QRFQFTlong}.

\section{Experimental proposal}

A plethora of experimental platforms based on Bose--Einstein condensates of ultracold atoms~\cite{Steinhauer:2015saa,Eckel:2017uqx,MunozdeNova:2018fxv,Viermann2022Nature,Tajik:2023,Martirosyan_2025} and on quantum fluids \cite{Falque:2023ctx,Delhom_2024,jarema2025informationquantumfieldtheory} have been developed in recent years within the analog-gravity program~\cite{Barcelo:2026vlr}.~Regardless of their connection with gravitational effects, we argue that they constitute ``quantum field simulators'' capable of testing the physics of QRFs in QFT (see~\cite{Gooding_2024,perche2026bosepolaronsrelativisticunruhdewitt} for other proposals to explore RQI effects with these systems).

For instance, for a dilute gas of bosonic atoms well below the critical temperature and in the limit of many atoms, the expectation value of the Bose field operator $\hat\Psi(\vec x)$---the condensate wavefunction---obeys the Gross--Pitaevskii equation (see e.g.~\cite{Barcelo:2026vlr,Delhom:2025dcj}).~The polar decomposition $\hat\Psi(\vec x)=:\sqrt{\hat n(\vec x)}\,e^{i\hat\theta(\vec x)}$ formally defines the density and phase operators $\hat n(\vec x)$ and $\hat\theta(\vec x)$, whose fluctuations are defined by expanding around their mean-field values, $\hat n(\vec x)=n_0(\vec x)\, \hat{ \mathbb{I}}+\delta\hat n(\vec x)$ and $\hat\theta(\vec x)=\theta_0(\vec x)\, \hat{ \mathbb{I}}+\delta\hat\theta(\vec x)$;~linearizing the equations of motion yields the dynamics of $\delta\hat n$ and $\delta\hat\theta$.

It is by now well understood~\cite{Unruh:1980cg,Barcelo:2026vlr,Delhom:2025dcj} that, in the hydrodynamical approximation---i.e., on scales much larger than the healing length $\xi\sim\hbar/(m c_{\rm s})$, with $m$ the atomic mass and $c_{\rm s}$ the speed of sound---$\delta\hat n(\vec x)$ and $\delta\hat\theta(\vec x)$ are described by a massless scalar field obeying a covariant Klein--Gordon equation on an effective curved spacetime.~Up to normalization constants, $\delta\hat\theta(\vec x)$ plays the role of the scalar field $\hat\Phi(\vec x)$, and $\delta\hat n(\vec x)$ that of its conjugate momentum $\hat\Pi(\vec x)$.~In particular, the shift symmetry $\hat\Phi\mapsto\hat\Phi+\lambda\,\hat{\mathbb{I}}$ corresponds to the global $U(1)$ symmetry $\hat\Psi\mapsto e^{i\lambda}\hat\Psi$ of the atomic gas.~The effective spacetime bears no relation to the one we live in and does not satisfy Einstein's equations;~its metric is determined by the mean density and flow velocity of the condensate, and it is flat when the condensate is homogeneous, stationary, and at rest.

Current experiments can modify not only the effective curvature~\cite{Viermann2022Nature} but also the effective spatial dimensionality \cite{Tajik:2023,Viermann2022Nature,Martirosyan_2025}, by confining the condensate tightly enough in one or two directions to freeze its transverse excitations.~The reduced theory inherits the $U(1)$ symmetry and therefore remains shift-invariant;~in $1+1$ dimensions, however, it takes the form of a minimally coupled Klein--Gordon equation only if $n_0/c_{\rm s}$ is position-independent (see, e.g.,~\cite{Anderson_2013}).~The platform of~\cite{Viermann2022Nature} has reproduced a $(2+1)$-dimensional scalar field theory with cold atoms, while~\cite{Steinhauer:2015saa,MunozdeNova:2018fxv,Tajik:2023} have reproduced $(1+1)$-dimensional configurations.

In equilibrium, the state of the effective scalar field theory is, to a good approximation, a thermal Gaussian state; the validity of the Gaussian approximation can be tested explicitly~\cite{Tajik:2023}.~Gaussianity implies that the state is fully determined by its first moments, which vanish in equilibrium, and its covariance matrix, i.e., the two-point correlation functions of $\delta\hat\theta(\vec x)$ and $\delta\hat n(\vec x)$, including the field--momentum cross-correlations.~Such tomographic reconstructions have been demonstrated with cold atoms in $1+1$~\cite{Tajik:2023} and $2+1$ dimensions~\cite{Viermann2022Nature}, and with quantum fluids in $2+1$ dimensions~\cite{Boulier:2014} (see also~\cite{Claude_2022,Agullo:2024lry}), showing that the experimental capabilities needed to test QRF-related physics in QFT already exist.

The core observation is that, within these approximations, the covariance matrix is all one needs to compute the states of relational subsystems, and the physical quantities derived from them.~As described above, this only requires smearing the measured two-point functions with the appropriate mode functions ($f_S-(\bar f_S/\bar f_R)f_R$ for a relational field), a post-processing of the experimental data technically analogous to the Fourier transform of two-point functions routinely performed (see e.g.~\cite{Viermann2022Nature}).

Because of the shift symmetry, $\delta\hat\theta(\vec x)$ can only be measured relative to a phase origin, fixed by the imaging system, which plays the role of an external reference.~If the measured data were processed as described above, which amounts to implementing the relativization map, one would obtain quantities defined relative to an internal, localized QRF.~Such quantities do not depend on the origin, and make the external reference a mere auxiliary tool with no impact on the relational predictions associated with internal QRFs.~For instance, the mutual information reported in \cite{Tajik:2023} refers to the field relative to the external origin, made meaningful by a tunnel coupling that pins the relative phase during the preparation of the state.~The data could be reanalyzed relative to localized internal frames, providing a direct test of the frame dependence predicted here.

An additional word of caution concerns infrared divergences.~For a massless field, correlation functions of smeared field operators $\hat\Phi_f$ with $\bar f\neq0$ diverge in $1+1$ dimensions, both in the vacuum and at finite temperature, and, at finite temperature, also in $2+1$ dimensions, whereas they remain finite whenever $\bar f=0$.~In the laboratory, all correlation functions are finite, owing to effective infrared cut-offs built into experimental platforms, such as the finite size of the system.~The resulting values of the otherwise divergent quantities, however, depend on the cut-off.~Shift-invariant quantities, in contrast, have a finite limit as the cut-off is removed, and are insensitive to it up to finite-size corrections that become negligible when  the effective cut-off is much larger than the region where the relevant smearing functions are supported.

Finally, it is worth stressing that an experimental realization of the physics discussed in this article would provide more than a test of our calculations.~Several aspects of real platforms are difficult to model from first principles, such as the precise POVMs implemented by the detection apparatus, or deviations from the idealized regime in which the analogy with a Klein--Gordon field emerges---e.g., dispersive effects at scales approaching the healing length, interactions between phonons, inhomogeneities, or the finite size of the system.~Since the shift symmetry descends from the exact $U(1)$ symmetry of the atomic gas, rather than from the approximations underlying the Klein--Gordon description, the relational framework remains meaningful in these regimes.~Quantum field simulators can therefore inform us about the physics of QRFs precisely where our theoretical modeling begins to fail.

\section{Final comments}

We have addressed the questions posed in the introduction for a massless scalar field with a shift symmetry.~(i) Localized modes of the field itself can serve as QRFs;~equipped with a covariant POVM, they define, through the relativization map, shift-invariant relational observables and relational subsystems, without recourse to external frames or to other fields.~(ii) Such frames differ in essential ways from their counterparts in systems with finitely many degrees of freedom:~in physically relevant states they are necessarily mixed, they are generically correlated with the systems they describe, and relational quantities remain finite in situations where their non-relational counterparts are infrared divergent.~(iii) The frame dependence of subsystems, entropies, and correlations, familiar from QRFs in quantum mechanics, extends to QFT, where it acquires a characteristic dependence on the separation between system and frame.~This dependence leads to a rich behavior of the mutual information and other correlation measures as functions of the location of the frame and of the temperature of the field.~(iv) These effects are accessible with existing simulators of quantum fields.

The use of a physical system with indefinite particle number as a phase reference goes back to Aharonov and Susskind's analysis of superselection rules~\cite{Aharonov:1967zza}.~A Bose--Einstein condensate was later proposed in this role \cite{Dowling:2006}---specifically, 
the condensate's global phase, conjugate to the total particle number.~Although treated quantum mechanically, such a reference approaches an idealized phase reference in the limit of large particle number, and its behavior admits an effective classical description~\cite{Loveridge:2017pcv}.~In our proposal, the global phase of the condensate plays at most the role of the auxiliary external reference discussed above.~The present work departs from this picture in three respects.~First, the frame is a localized mode of the phonon field itself---rather than a separate system---so that frame and system are degrees of freedom of the same QFT, correlated through its state.~Second, the frame is genuinely quantum:~no large-reference limit is invoked, its fluctuations and its correlations with the rest of the field are fully taken into account, and, in physically relevant states, it is necessarily in a mixed state.~Third, the predictions are testable with existing technology, since all quantities of interest follow, by post-processing, from covariance matrices whose tomographic reconstruction has already been demonstrated~\cite{Tajik:2023,Viermann2022Nature}.

Several extensions are natural.~The shift symmetry was chosen as the simplest analogue of gauge transformations, and the tools developed here can be extended to electrodynamics, where the gauge constraint introduces the additional restrictions on physical states discussed above.~Analog platforms also give access to curved and time-dependent effective spacetimes, where the frame dependence of subsystems and correlations could be explored in the presence of horizons and particle creation.~Quantum simulators based on quantum gases and fluids thus offer a concrete arena in which the relational nature of physical quantities in QFT can be put to experimental test.

\begin{center}
\textbf{Note added}
\end{center}
During the completion of this work and its companion article~\cite{QRFQFTlong}, we became aware that two other groups have independently investigated ideas that partially overlap with, and complement, the analysis presented here:~\cite{FL_QRFQFT}, which also appeared in today's listing, and the forthcoming work \cite{Giacomini:RegionalQRF}.~We look forward to comparing our works.

\begin{center}
\textbf{Acknowledgments}
\end{center}

The authors have benefited from discussions with E.~Bianchi, L.-Q.~Chen, A. Delhom, A.-C.~de la Hamette, L.~Freidel, F.~Giacomini, P.~A.~H\"ohn, J.~Kirklin, G.~Lo Giudice, L.~Loveridge, L.~Marchetti, E.~Mart\'in-Mart\'inez, E.~Panella, T.~R.~Perche, P.~Ribes-Metidieri, C.~Rovelli, and M.~Varadarajan.~This work was supported in part by the NSF grants PHY-2409402 and by the Hearne Institute of theoretical physics at Louisiana State University.~This project/publication was also made possible through the support of the COST Action CA23115 and \href{https://withoutspacetime.org}{WOST, WithOut SpaceTime project}, led by the Center for Spacetime and the Quantum (CSTQ), and supported by Grant ID\# 63683 from the John Templeton Foundation.~The opinions expressed in this project/publication are those of the author(s) and do not necessarily reflect the views of the John Templeton Foundation.~I.A.~was partially supported by the visiting fellow program of Perimeter Institute for Theoretical Physics.~Research at Perimeter Institute is supported
by the Government of Canada through the Department of Innovation, Science, and Economic
Development, and by the Province of Ontario through the Ministry of Colleges and Universities.~F.M.M.~acknowledges the hospitality of the Okinawa Institute of Science and Technology and the Perimeter Institute for Theoretical Physics, where part of this work was carried out. 

\bibliography{QFTQRF}
\end{document}